
%
%
\magnification=\magstep1
%
%
%
%
%
\parskip=2pt plus1pt
\parindent=2em
\baselineskip=13pt
%
%
\catcode`\@=11
\topskip=10pt plus 10pt
\r@ggedbottomtrue
\catcode`\@=12
%
%
\font\sslarge=cmss17
%
%
\font\eightrm=cmr8
\font\sixrm=cmr6

\font\eighti=cmmi8
\font\sixi=cmmi6
\skewchar\eighti='177 \skewchar\sixi='177

\font\eightsy=cmsy8
\font\sixsy=cmsy6
\skewchar\eightsy='60 \skewchar\sixsy='60

\font\eightbf=cmbx8
\font\sixbf=cmbx6

\font\eighttt=cmtt8

\hyphenchar\tentt=-1 
\hyphenchar\eighttt=-1

\font\eightsl=cmsl8

\font\eightit=cmti8

\newskip\ttglue

\def\eightpoint{\def\rm{\fam0\eightrm}%
  \textfont0=\eightrm \scriptfont0=\sixrm \scriptscriptfont0=\fiverm
  \textfont1=\eighti \scriptfont1=\sixi \scriptscriptfont1=\fivei
  \textfont2=\eightsy \scriptfont2=\sixsy \scriptscriptfont2=\fivesy
  \textfont3=\tenex \scriptfont3=\tenex \scriptscriptfont3=\tenex
  \def\it{\fam\itfam\eightit}%
  \textfont\itfam=\eightit
  \def\sl{\fam\slfam\eightsl}%
  \textfont\slfam=\eightsl
  \def\bf{\fam\bffam\eightbf}%
  \textfont\bffam=\eightbf \scriptfont\bffam=\sixbf
   \scriptscriptfont\bffam=\fivebf
  \def\tt{\fam\ttfam\eighttt}%
  \textfont\ttfam=\eighttt
  \tt \ttglue=.5em plus.25em minus.15em
  \normalbaselineskip=9pt
  \let\sc=\sixrm
  \let\big=\eightbig
  \normalbaselines\rm}
%
%
\newcount\sectionnum
\newcount\subsectionnum
\newcount\eqnum
\global\sectionnum=0
\global\subsectionnum=0
\global\eqnum=0
\def\section#1{\global\advance\sectionnum by 1\global\subsectionnum=0
    \goodbreak\bigskip\centerline{\the\sectionnum. #1}
    \nobreak\medskip}
\def\subsection#1{\global\advance\subsectionnum by 1 \goodbreak\medskip
    \centerline{\the\sectionnum.\the\subsectionnum. \it #1}
    \nobreak\smallskip}
%
%
\def\puteqnum{\global\advance\eqnum by 1 \eqno{(\the\eqnum)}}
%
%
\def\abstract#1{\centerline{\sl ABSTRACT}\smallskip
    {\narrower\eightpoint\par #1\par}}
%
%
\def\subjectheadings#1{\smallskip{\narrower\eightpoint\par\hangindent 8em
      \noindent\hskip 8em
      \llap{\it Subject headings: }#1\par}\bigskip}
%
%
\footline={\ifnum\pageno>1{\hss\tenrm\folio\hss}
           \else{\ifnum\pageno<-2{\hss\tenrm\folio\hss}
                 \else{\hfil}
                 \fi}
           \fi}
%
%
%
%
%
%
\def\smallup#1{\raise 1.0ex\hbox{\sixrm #1}}
\def\smallvfootnote#1#2{\vfootnote{{\eightrm\raise 1.0ex\hbox{\sixrm#1}}}
       {\baselineskip=10pt\eightrm #2}}
%
%
\def\figpar{\par\noindent\hangindent=1.5em}
\def\figcapt#1{\filbreak\smallskip\figpar{\bf Figure #1}\ --\ }
\def\startfigcapt{\vfill\eject
     \centerline{\bf Figure Captions}\nobreak\medskip}
\def\endfigcapt{\vfill\eject}
%
%

%
%

%
%

\def\ltsima{$\; \buildrel < \over \sim \;$}
\def\ltsim{\lower.5ex\hbox{\ltsima}}
\def\gtsima{$\; \buildrel > \over \sim \;$}
\def\gtsim{\lower.5ex\hbox{\gtsima}}

%
%

\def \Msun      {{\rm\,M}_{\odot}}

\def \kpc       {{\rm\,kpc}}

%
%
\def\startreferences{\vfill\eject\centerline{\bf References}}
\def\endreferences{\vfill\eject}
%
%
\def\refpar{\par\noindent\hangindent=1.5em\frenchspacing}
\def\bref{\refpar}
\def\jref#1;#2;#3;#4;#5;#6.{\refpar #1 #2, #3 #4, #5}
\def\jrefabbrev#1;#2;#3;#4;#5;#6;#7.{\refpar #1 #2, #3 #4, #5 (#7)}
\def\jrefprivcomm#1;#2;#3;#4;#5;#6.{\refpar #1 #2, private
       communication}
\def\jrefinprep#1;#2;#3;#4;#5;#6.{\refpar #1 #2, in preparation}
\def\jrefpreprint#1;#2;#3;#4;#5;#6.{\refpar #1 #2, preprint}
\def\jrefsub#1;#2;#3;#4;#5;#6.{\refpar #1 #2, #3, submitted}
\def\jrefpress#1;#2;#3;#4;#5;#6.{\refpar #1 #2, #3, in press}
%
%

\def\apj{ApJ}
\def\apjl{ApJ}
\def\apjs{ApJS}

\def\cip{Comput. Phys.}

\def\mnras{MNRAS}
\def\nature{Nature}

\def\physrevlett{Phys. Rev. Lett.}

%
%
\def\xten#1{\times 10^{#1}}
\def\etal{et al{.}\ }

\def\eg{e{.}g{.},\ }

\def\numsym{\raise 0.4em\hbox{${\scriptstyle\#}$}}

%
%
%
\def\bvp{v_{b,pairwise}}
\def\bvc{v_{b,cluster}}
%
%
\def\author#1{\smallskip\centerline{#1}}
\def\institution#1{\centerline{\eightpoint #1}}
\def\andauthor#1{\smallskip\centerline{and}\smallskip\centerline{#1}}

\bigskip
\centerline{\sslarge Galaxy Tracers and Velocity Bias}
\medskip
\author{F J Summers\smallup{1} and Marc Davis\smallup{2}}
\institution{Astronomy Department, University of California,
   Berkeley, CA 94720}
\andauthor{August E. Evrard\smallup{3}}
\institution{Physics Department, University of Michigan,
   Ann Arbor, MI 48103}
\smallvfootnote{1}{Present address: Princeton University Observatory,
   Peyton Hall, Princeton, NJ 08544;\hfil\break\hbox{ }\qquad
   summers@astro.princeton.edu}
\smallvfootnote{2}{Also Physics Department, University of California at
   Berkeley; marc@coma.berkeley.edu}
\smallvfootnote{3}{evrard@pablo.physics.lsa.umich.edu}

%
%
\bigskip\bigskip
%
%

\abstract{
This paper examines several methods of tracing galaxies in N-body
simulations and the effects of these methods on the derived galaxy
statistics. Special attention is paid to the phenomenon of velocity
bias, the idea that the velocities of galaxies may be
systematically different from that of the mass distribution. Using two
simulations with identical initial conditions, one following a single
dark matter particle fluid and the other following two particle fluids
of dark matter and baryons, both collisionless and collisional methods
of tracing galaxies are compared to one another and against a set of
idealized criteria. None of the collisionless methods proves
satisfactory, including an elaborate scheme developed here to circumvent
previously known problems. The main problem is that galactic
overdensities are both secularly and impulsively disrupted while
orbiting in cluster potentials. With dissipation, the baryonic tracers
have much higher density contrasts and much smaller cross sections,
allowing them to remain distinct within the cluster potential. The
question remains whether the incomplete physical model, especially the
expected conversion from a collisional gas to a collisionless
stellar fluid, introduces systematic biases. Statistical measures
determined from simulations can vary significantly based solely on
the galaxy tracing method utilized. The two point correlation function
differs most on cluster scales (less than 1 Mpc) with generally
good agreement on larger scales (except for one systematically biased
method). Pairwise velocity dispersions show less uniformity on all
scales addressed here. All tracing methods show a velocity bias to
varying degrees, but the predictions are not firm: either the tracing
method is not robust or the statistical significance has not been
demonstrated. Though theoretical arguments suggest that a mild velocity
bias should exist, simulation results are not yet conclusive.
}

%
%
\subjectheadings{cosmology: large-scale structure of universe ---
   galaxies: clustering --- galaxies: formation --- methods: numerical}

%
%
\bigskip\bigskip

N-body simulations of large scale structure have sometimes
been referred to as `experimental cosmology'. Indeed, at the heart of
these efforts is the desire to provide stringent tests of various
theories of structure development. Much progress has been made in
determining which theories produce structure similar to what is
observed, but a major obstacle to this progress is the identification of
where galaxies would reside in a numerical simulation. Accurate
identification of galaxy positions and velocities within the computer
models is required in order to provide reliable estimates of important
structure measurements such as the correlation function and the velocity
dispersion of galaxies. Tracing galaxies is also essential for
following the processes of galaxy, group, and cluster formation. As
the results derived may support or reject a cosmogony, the various
methods for tracing galaxies must be evaluated carefully.

During the last decade, particle simulations became a standard tool
for evaluating cosmological theories. Simulations
were influential in shifting focus away
from the hot dark matter scenario (White, Frenk, \& Davis 1983) and
toward the cold dark matter (CDM) scenario (Davis \etal 1985). Galaxy
redshift surveys are routinely compared against computational
realizations of cosmological theories in order to assess the viability
of the theories (\eg Efstathiou, Sutherland, \& Maddox 1990, Vogeley
\etal 1992, Fisher \etal 1995).  Simulations have also played a
significant role in the recent attention on the mixed dark matter
scenario (Davis, Summers, \& Schlegel 1992, Klypin \etal 1993). N-body
calculations are now an important method for determining the
predictions of a cosmological theory in the non-linear regime.

Unfortunately, comparison of a computer model with
observations involves several caveats, restrictions, and subtleties of
interpretation that limit the range of conclusions that can be made. One
of the most important to note is that by no stretch of the imagination
does one form `galaxies' in current cosmological simulations: the physical
model and dynamic range are inadequate to follow any but the crudest of
details of such complex and rich structures. What
one hopes to do is find the likely sites of galaxy formation and trace
their evolution. It is the statistics of these galaxy tracers that are
used to evaluate the theories and thus, it is of paramount importance
that the galaxy tracer population be as reliable as possible.

The importance of the galaxy tracer population has been illustrated
quite vividly. In early simulations of the CDM theory,
Melott \etal (1983) found that the correlation function of the mass could
replicate the observed correlation function of galaxies. Using more detailed
simulations and an algorithm for identifying galaxy tracers,
Davis \etal (1985) found that galaxies had to be more correlated than
the dark matter to get an acceptable fit to the data. The excess
correlation strength was termed `biasing' and their results showed it
to be a rather strong effect. Couchman and Carlberg (1992, hereafter
CC) also modelled a CDM universe, but, using a different
galaxy tracing method, reached a third conclusion.
Previously, Carlberg, Couchman, \& Thomas (1990) had proposed that the
velocity field of galaxies could be systematically lower than that of
the dark matter and dubbed this new effect velocity bias in order to
differentiate it from the correlation bias. CC found a significant
velocity bias such that the data could be consistent with a CDM
model where the galaxies were actually slightly less correlated
than the dark matter (a mild anti-bias). These claims have sparked a lot
of discussion because CDM with large biasing is at odds with several
measures of large scale structure such as the APM galaxy distribution
(Efstathiou \etal 1990) and the COBE observations (Efstathiou, Bond,
\& White 1992). From a computational scientist's point of view, one
should be very uncomfortable with such a large impact on one's
conclusions based heavily on a galaxy tracer algorithm.

Two groups have recently reported a notable improvement in galaxy
tracing. The work in our group (Summers 1993,
Evrard, Summers, \& Davis 1994, hereafter Paper I) will be referred to
as SDE and the other group by their author list, KHW (Katz, Hernquist,
\& Weinberg 1992).  The simulations of these researchers employed two
particle fluids, one of dark matter and one of baryons (also called
gas), and enough physics to follow the collapse of the luminous parts
of galaxies directly. For the first time, objects of galactic
densities were found in cosmological simulations and
reasonable populations of galaxy-like objects could be defined, albeit
at considerable computational expense.  Interestingly, on the question
of velocity bias, SDE found results mostly in accord with CC,
while KHW found no evidence for velocity bias.

Some aspects of velocity bias need to be clarified. As Carlberg (1994)
points out, there are two forms of velocity bias, one for the global
pairwise velocity dispersions, $\bvp$, and one for the velocity
dispersions within clusters, $\bvc$ (Carlberg refers to these as the
two particle and single particle velocity biases, respectively).  The
first compares the pairwise velocity dispersion of galaxies, or galaxy
tracers, to that of the mass distribution. This pairwise velocity bias
can be a function of scale and is usually compared at 1 Mpc
$$
\bvp={\sigma_{p,g}\over \sigma_{p,\rho}}\Big|_{1\ \rm Mpc}
\puteqnum
$$
where $\sigma_p$ represents a one dimensional pairwise velocity
dispersion and the subscripts $g$ and $\rho$ denote the galaxies and
the mass, respectively.  When considering a single cluster of
galaxies, a similar ratio for the cluster velocity bias,
$$
\bvc={\sigma_g\over\sigma_\rho}\Big|_{\rm within\ cluster}\ \ ,
\puteqnum
$$
indicates the degree to which the galaxy velocities sample the full
cluster potential. Carlberg (1994) shows analytically how $\bvc$ may
be understood, but requires N-body simulations to estimate its value.
Because the velocity dispersions in large clusters can dominate the
pairwise statistic (Gelb \& Bertschinger 1994), the two forms are not
independent and may have similar values, a confusing point for
researchers. The global statistic, $\bvp$, has been considered by many
simulations (Carlberg \& Couchman 1989, Carlberg, Couchman, and Thomas
1990, Couchman \& Carlberg 1992, Katz \etal 1992, Cen \& Ostriker 1992,
Evrard \etal 1994) and may be compared to the observed dispersion of
galaxies in the nearby universe (Davis \& Peebles 1983). The cluster
statistic, $\bvc$, is important for dynamical estimates of cluster
masses and comparison to x-ray observations (Carlberg \& Dubinski
1991, Lubin \& Bahcall 1993, Katz \& White 1994, Carlberg 1994,
Frenk \etal 1995).

The current status of velocity bias is somewhat confused, in no small
part because making reliable estimates of velocity bias from
simulations is quite difficult. One needs a simulation with large
enough dynamic range to cover from galaxy scales ($\sim$10 kpc) up to
scales that cover a statistically fair sample of the universe
($\sim$100 Mpc). Also, because this effect may be concentrated in
clusters, the ability of the galaxy tracing method to resolve cluster
regions is paramount. If dynamical friction is the cause (Carlberg
\etal 1990), then the galaxy tracers must be much denser than the
surrounding medium in order to test it.  Measurements of both forms of
velocity bias have ranged from 0.3 to 1.0, but the effect of differing
galaxy tracer algorithms is not known.  With such a large variation in
the conclusions, it is a fair question to ask whether any of the
estimates are correct.

This paper evaluates and compares several galaxy tracing methods and
examines the results derived from them. Criteria of a what constitutes
a `good' galaxy tracer is developed and applied to the different
tracing methods.  Using two simulations with the same initial
conditions, one with dark matter only and the other employing both dark
matter and gas, the various algorithms are considered on an even
footing. An understanding of each method's strengths and limitations is
required for applying them properly. We also test whether the extra
complexity and cost of the two fluid simulations are necessary. In order
to get an idea of the consequences when comparing to real data, the
correlation and velocity statistics of each population will be examined
and contrasted. We will then be able to make better statements
on the issue of velocity bias.

\section{Criteria and Methods}

In order to evaluate the various galaxy tracing methods, it is useful to
have a set of ideal criteria that can serve as a basis for judgement. In
this respect, the main concern is for finding and following the
positions and velocities of the galaxy tracers since these are the basic
properties that are used in comparison against real data. Other
characteristics such as luminosity, morphology, and internal dynamics
provide the details of galaxy formation, but are not strictly necessary
for just tracing galaxies. Additionally, the statistical measures of a
good tracer population should reflect the assumed cosmology and are not
inherent in the tracer algorithms. Thus, finding the expected
correlation function does not validate the tracer algorithm. The
limitations and ramifications when interpreting these statistics will be
discussed in a later section.

The first thing one requires is that the tracers be well defined. The
positions and velocities should be uniquely specified and the dependence
on the parameters of the algorithm should be weak.  Some classification
of the galaxy tracers, usually according to mass, is required for one to
be able to compare an observed galaxy population against the correct
counterpart in the simulation. Often these comparisons will involve
further assumptions, such as a mass to light ratio, but the initial
basis for comparison must be an attribute of the tracers.  These are
minimum criteria for an identifiable population.

Further, because one is ostensibly studying galaxies, other criteria are
needed. The evolution of an individual tracer should be coherent in that
it can be easily tracked from one simulation output to the next.  The
merging of tracers as well as the possible disruption of a tracer should
be addressed by the method. For robust statistical measures, the tracers
must be identifiable in both field and cluster environments. And
finally, there is a subjective measure of the plausibility of the tracer
selection algorithm for picking out and tracking the likely sites of
galaxy formation. Although a plausibility criterion sounds open ended,
the emphasis is that the identification and evolution of galaxy tracers
be shown to follow a reasonable path and that neither becomes corrupted
by numerical artifacts. Satisfying these criteria will not ensure a
correct population, but will go a long way toward enhancing one's
confidence in the selection method. In most cases, these criteria are
useful in identifying the limitations of a population.

The best way to compare the various galaxy tracing methods is to apply
each of them to the same simulation. Each method will examine the same
particle configurations and will determine what structures it can find.
To explore both collisionless and dissipational methods of tracing
galaxies, we will use two simulations evolved with the P3MSPH code
(Summers 1993, Evrard 1988) from the same initial conditions; the first
uses a single collisionless dark matter fluid and the second follows
two fluids: dark matter and a dissipational baryonic component. The
Two Fluid run was described in detail in Paper I and the DM run is the
single fluid version. Important parameters of the simulations include a
comoving cubical size of 16 Mpc along a side, a particle count of
262,144 particles per species, and an effective resolution of
$\sim 20/(1+z)\kpc$. The initial conditions use a CDM power spectrum and
are constrained to produce a mass concentration characteristic of
a poor cluster of galaxies in the
center of the region. The DM run has a timestep of 4.3 Myr and a
particle mass of $1.08\xten{9}\Msun$. The Two Fluid run uses a timestep
of 6.1 Myr, has particle masses of $9.72\xten{8}\Msun$ and $1.08\xten{8}$
for the dark matter and baryonic particles, respectively, and models
thermal pressure, shock heating, and radiative cooling via smoothed
particle hydrodynamics (SPH) for the baryonic species. The simulations
have sufficient time, length, and mass resolution to follow galaxy
formation and can address both field and cluster regions.

Figure 1 shows the major structures to be examined in the simulation at
the final time. The left plot shows the dark matter in the entire
computational volume and the right plot details the central region, one
tenth of the box length on a side. For clarity, the full box plot shows
only one fourth of the particles in the simulation. The simulation forms
a poor cluster, $3.5\xten{13}\Msun$, in the center, a group of galaxies
above and right of center, and a variety of filamentary structure
typical of hierarchical simulations. The central region is dominated by
a massive halo with little apparent substructure in the dark matter.
This figure will be used as a reference for the tracer algorithms.

\section{Collisionless Algorithms}

In cosmological theories where the universe is dominated by weakly
interacting dark matter, simulations which use a single species of
collisionless particles interacting only via gravity are appropriate
and offer several advantages.  Since gravity is the dominant force on
large scales, the galaxy distribution should, for the most part, trace
the dark matter distribution. Implementing gravitational physics in
simulations is well studied and the codes are robust and optimized.
The caveat is that one cannot state for certain that the baryons in
the universe will closely follow the dark matter and one knows that
gas physics plays an important role in galaxy formation.
Hence, a large amount of leeway is provided by schemes which bias the
galaxy distribution relative to the dark matter and the constraints
one can place on cosmological models are not as stringent as may have
been hoped. Still, the large majority of cosmological simulations are
collisionless ones.

We will discuss four collisionless galaxy tracing algorithms. The
first two, which we refer to as halos and peaks, are well known in the
literature and our discussion will be brief. The third is somewhat
newer, having been suggested by Couchman and Carlberg (1992), but was
not tested in great detail. The fourth is a new algorithm developed
here.

\subsection{Halos}

The most natural method for choosing where galaxies are likely to form is
to examine the evolved density field. One identifies dense regions in
the dark matter distribution as the dark matter halos of galaxies and
of clusters of galaxies. This method is both simple and straightforward.

In practice, one utilizes a grouping algorithm to identify groups of
particles as galaxy halos. We will use the friends of friends
algorithm (FOF) which gathers into a group all particles separated by
less than a specified linking length, $\eta$. In the usual custom, the
linking length is specified in units of the mean interparticle
separation, taken to be $L/N$ for a cube of side length $L$ containing
$N^3$ particles.  It is usually more informative to give an
overdensity constraint for the groups. We shall quote a value
$\delta_{min}$ that specifies the minimum overdensity contour of a
group and is defined by two particles within a sphere of linking
length radius:
$$
\delta_{min} = {2 \Omega_{sim} \over {4\over3} \pi \eta^3}
\puteqnum
$$
where this formula applies for equal mass particles and the $\Omega_{sim}$
factor represents the overdensity of our simulation volume compared to the
background universe (due to constrained initial conditions).
Be aware that other researchers may have
used a different estimate of the overdensity, namely $\Omega_{sim}/\eta^3$
(\eg Davis \etal 1985).  Additionally, $\delta_{min}$ is the minimum
contour level and the averaged overdensity of the group is generally
many times larger.  The other parameter that needs to be specified is
the minimum number of particles required to make a group, $N_{min}$.

The main problem with the halos method has been recognized from the
start (Davis \etal 1985). Simply put, when galaxy sized halos merge to
form a cluster, the substructure in the cluster halo is erased in about
a crossing time and one cannot continue to track individual galaxy halos
within the cluster halo. In terms of the criteria outlined above, the
method fails to resolve both field and cluster regions.

An example of the method is shown in Figure 2. We have applied the FOF
algorithm to the simulation with $\delta_{min}=300$, $N_{min}=30$ and
plot the halos in the same views as shown in Figure 1. The method does
a good job in regions where a single dark matter halo can be assumed
to contain only one galaxy, but fails to find substructure in the
central cluster region where many galaxies would be expected. Figure 3
illustrates the merging process by plotting the particles in the
largest halo at an early output and at the end of the simulation. A
collisionless fluid quickly erases visual substructure.

\subsection{Peaks}

The peaks method arises naturally from the theory of collapse of
density perturbations (Bardeen \etal 1986) and can circumvent most of
the merging problems of the halo algorithm. Instead of looking for
galaxy tracers in the evolved density field, one examines the initial
density field. The procedure, in rough terms, is to smooth the initial
density field on a desired filtering scale, to identify peaks in the
smoothed field above a given peak height threshold as likely sites for
galaxy formation, and to tag a single particle near the center of each
peak as a galaxy tracer of that peak.  Two versions of the method have
been developed: the peak tracer method, when one can resolve galaxy
scale peaks (Davis \etal 1985), and the peak/background split, when
galaxy scales are not well resolved (White \etal 1987). Park (1991)
showed that the results of the two methods are similar.

In terms of the definite criteria described in Section 2, the peaks
method satisfies, or can be made to satisfy, all of the concerns.
Because one is using a single particle as a tracer of a galaxy, there
are no questions about being well defined and coherent. One objection
may be that a prescription for merging of the peak tracers must be
devised, but it does not raise fatal difficulties (White \etal 1987).
Additionally, the method presumes no disruption of galaxies. A
particular advantage of this method is that by adjusting the filtering
scale and the threshold peak value, one can match the number density
of peaks to the number density of the galactic objects being studied
at the outset. This should not be considered fine tuning, but rather a
requirement that one use the correct simulated population to compare
with the population in the data.

The problems of the peaks method lie in its correlation with the
evolved density field. Although it is a reasonable expectation, there
is no guarantee that galaxies will form only from the highest peaks
apparent in the linear density field. Further, phase space mixing and tidal
distortions are prominent processes in a collisionless fluid (Moore,
Katz, \& Lake 1995) and it is not clear that a
single particle will accurately trace the evolution of a peak. Katz,
Quinn, \& Gelb (1993) have addressed the correspondence of linear peak
tracers and the evolved density field by comparing
the results of the peaks algorithm and the halo algorithm on the same
simulation. Their results confirm that both of the above concerns are
valid: halos do not necessarily arise from high peaks and high peak
particles do not necessarily end up in halos. The high frequency of
these occurrences, of order 30\%, undermines one's faith in the peaks
method. Similar conclusions on the collapse of density peaks were also
found by Gelb \& Bertschinger (1994) and from a theoretical standpoint
by Bertschinger \& Jain (1994).

\subsection{Couchman and Carlberg's Method}

Another method for avoiding the merging problems of the halo method
was suggested in the CC paper. Their idea was to tag particles in
galaxy sized halos before a cluster forms. After the cluster merges,
one can use those tagged particles to identify where the galaxies
would be inside of the cluster halo. One might think of this procedure
as a two stage halo algorithm: identify halos at an early output and
then identify groups from those halo particles at a late output. In
this manner, one might resolve both cluster and field regions. The
idea is a good one, but CC did not adequately show that it works in
practice.

The method as described in CC is only a little more complex than the
halo method. At a chosen redshift, $z_{tag}$, one employs a grouping
algorithm with parameters $\delta_{tag}$ and $N_{tag}$ and tags all
particles in those groups. At the final output of the simulation, one
performs a grouping operation on only the tagged particles using
parameters $\delta_{fin}$ and $N_{fin}$. CC then do a mass range cut of
these final groups, accepting only groups with particle numbers between
$N_{low}$ and $N_{high}$. The groups that remain are considered to be
galaxy tracers.

There are several problems with this implementation. First, the method
allows for only one epoch of galaxy tagging, while the collapse of
density peaks is considered to be a continuous process. Galaxy halos
must have collapsed by $z_{tag}$ in order to be included in the tracer
population.  The mass cut applied at the end is ostensibly to select a
given mass range of galaxy tracers, but the mass of these tracers is ill
defined because the method does not account for infall of material
between the time of tagging and the final time. The coherence of
the galaxy tracers is also unknown because there is no check that the
particles that end up in a final tracer were in proximity to one another
at the tagging epoch. And last, although CC state that $\delta_{tag}$
and $\delta_{fin}$ should ideally be related by corresponding to the
same physical density, in practice, they adjust the $\delta$ parameters
independently.

To explore the CC method in detail, we implemented it, with a couple
modifications, on our simulation. Our $z_{tag}$ had to be much earlier
than theirs, 6.6 versus 3, because our simulation is focused on
forming a cluster of galaxies. At redshift 3, much of the cluster halo
had already merged.  Similarly, our final output is at $z_{fin}=1$
instead of at $z_{fin}=0$, but note that the expansion factor between
tagging and final epochs is similar, 3.6 versus 4. With 36 times
higher mass resolution, we can implement a constant physical density
constraint on $\delta_{tag}$ and $\delta_{fin}$. Because we feel it is
not well motivated, the mass range cut was not implemented, though we
have studied what would happen if it were employed. Since the number
of CC tracers found was low compared to the other methods, $N_{tag}$
and $N_{fin}$ were set to 5; what we feel is the minimum number of
particles one might justify as a believable group.

The first conclusion is that using constant physical density to define
tracers does not work. The natural choice was to set $\delta_{tag}$ such as
to pick out suitable galaxy halos at the tagging epoch. However, since
the average density of the universe falls as $R^{-3}$, the corresponding
$\delta_{fin}$ becomes much too restrictive and very few galaxy tracers
are found (see Figure 4). The tracers are concentrated in the cluster
with very few in the field regions. Alternatively, if one sets
$\delta_{fin}$ by what would pick out a good field population at the
final time, then $\delta_{tag}$ is very low and one tags not only the
galaxy halos, but also the filaments and collapsing regions around them.
Too many particles are tagged and no structure is found within the final
cluster halo (see Figure 5). An intermediate density choice solves
neither problem. A constant physical density constraint in this method
does not help one resolve both field and cluster regions.

To produce a tracer population from the CC method, it was necessary to
choose $\delta_{tag}$ and $\delta_{fin}$ independently. The results
using $\delta_{tag}=300$ and $\delta_{fin}=750$ are shown in Figure 6
in the same plots as Figures 1 and 2. As a result of the single
tagging epoch, the distribution appears weak in the field regions
compared to the halos. There are approximately the same number of halo
and CC tracers found in the cluster region and the size of the largest
object is significantly reduced for the CC method.  However, upon
examining the velocity fields of the smaller CC tracers (5--10 particles)
in the cluster region, about half are incoherent and will
disperse within a crossing time of the cluster. These tracers appear to
be chance aggregations of particles originating in different halos at
$z_{tag}$. Such groupings are more common than one might expect because
the average density in the cluster region is well
above average and a much smaller fluctuation is required to create an
anomalous grouping. If one does not include these tracers, the cluster
region will be poorly sampled. Based on these results and the concerns
about the basic methodology expressed above, the CC algorithm fails
several of the galaxy tracer criteria.

An important point should be mentioned about CC's mass range cut. In
our simulation, CC galaxy tracers in the cluster region are dominated
by a large main halo surrounded by many much smaller ones.
Implementing a maximum particle number cutoff, $N_{high}$, will remove
the largest tracer and leave only a shell of tracers around the
central region. One essentially removes the potato and leaves only the
potato skin (see Figure 7).  Statistics of this population would not
be robust. Similar effects could have skewed the results presented by
CC.

In a recent paper, Carlberg (1994) has made some modifications to the
above method. He increases the minimum number of particles to 10 at
both epochs and checks for excessive velocity dispersions in the final
tracer groups to exclude unbound objects. The resulting method is a
small improvement, but it does not address the concerns about a single
tagging epoch or the disruption of the tracers while they orbit
through the cluster.  Results in these tests would only be slightly
improved over the CC method. In any case, his new method is much less
sophisticated than the most bound algorithm presented below.

\subsection{The Most Bound Algorithm}

While analyzing the CC method, it became apparent that the basic idea
was sound, but that several improvements could be made to the
implementation. To this end, we have attempted to create a collisionless
galaxy tracer algorithm that would withstand the {\it a priori\/}
objections to the CC implementation. The basic idea is that one would
like to catch each forming galaxy just after initial collapse when
correspondence between dark matter halos and galaxies is relatively
unambiguous. Since significant tidal dispersion of the outer regions
is expected during
collisionless merging, one should tag and follow only the core particles
of these nascent galaxies. Merging simulations have shown that particles
which are in the cores before merging tend to reside in the core of the
resulting object (Barnes 1989). At the final time, one can check which
cores have remained intact, merged, or been disrupted. We call this
method the most bound algorithm (hereafter MB).

The MB algorithm is rather complex. To allow for continuous collapse of
galaxies, we search for galaxy halos at 21 epochs evenly spaced in time
throughout the simulation. At each output we perform a friends of
friends grouping algorithm using a constant overdensity cutoff,
$\delta_{min}=300$. In addition, we remove from these
groups any particles that are gravitationally unbound and apply a
minimum particle number cutoff $N_{min}=30$. What we want to do is tag
the $N_{tag}=30$ most bound particles of each halo that has just
collapsed.

To identify the recently formed halos, each FOF group is checked to
see if it contains any tagged particles (i.e., has this group's core
already been tagged?). If yes, then the group is ignored. If no, then
the $N_{tag}$ most bound particles are tagged with the output time and
group identification. These groups are collectively referred to as the
tagging groups.  Checks are made to ensure that a few stray tagged
particles do not inhibit the tagging of a newly formed halo.

This procedure repeats at each output and builds up a population of
tagging groups, each with $N_{tag}$ particles. Then we identify where
every tagged particle is at the end of the simulation and, using only
these tagged particles, perform an FOF grouping with the same
$\delta_{min}$ and $N_{min,fin}=10$. Gravitationally unbound particles
could not be removed at this step because we are only using a subset
of the particles for grouping. This step identifies where the core
particles are at the end and identifies groups of core particles.
These final groups will be referred to as tracer groups.

Unfortunately, our search does not stop there. We must check the
coherence of these tracer groups and delete stray particles.  For each
tracer group, the particle tags are examined to find out which tagging
groups are represented. Strays are deleted by removing particles from
any tagging group represented by fewer than $f={1/3}$ of the original
$N_{tag}$ tagged particles. After stray deletion, the tracer groups are
patently coherent in that they contain combinations of $f\,N_{tag}=10$
or more particles that had been identified at higher redshift as being
together in phase space. A plot of the resulting 257 tracer groups is
shown in Figure 8.

Using the particle tags, one can probe the merging history.
If a tracer group contains particles from more than one tagging group,
it is considered to be a merger. A tagging group that is not represented
in any of the final tracer groups is considered to have been dispersed.
Note that it is also possible for a tagging group to be found in more
than one tracer group. We call these split groups.

The choices of parameters given above were influenced by several
considerations. The interval between output times is about 210 Myr, a
balance between the number of outputs we obtained from the simulation
and the collapse time of a resolvable halo. $N_{min}=N_{tag}$ was set
small to increase resolution of the method, but large enough that some
of the core particles could be scattered and still leave a believable
tracer.  $\delta_{min}$ corresponds to a nominal overdensity for a
perturbation that has just collapsed and virialized, the ones we are
interested in tagging. The minimum represented fraction,
$N_{min,fin}=fN_{tag}$, was based partly on plausibility (how small a
fragment would one believe is a {\it real} tracer) and partly on
empirical experience (smaller fragments showed incoherent velocity
patterns). Variations of the parameters produced consistent results,
though increasing $f$ by more than 50\% begins to undermine the method.

The MB method can be shown to satisfy all the definite criteria outlined
above. The identification algorithm is well defined and not too
sensitive on parameter choice. The minimum resolvable mass of the
procedure is determined by the parameter $N_{tag}$. Coherence, merging,
and disruption are dealt with explicitly. As Figure 8 shows, the field
and cluster populations are not unreasonable.  However, the indefinite
criterion, plausibility, requires further investigation.

The tapestry begins to unravel when one does an accounting of the tagging
groups. Over the 21 outputs, 402 groups were tagged. The 257 final tracer
groups includes 29 mergers of 2 to 11 tagging groups that account for 60
other tagging groups. 3 tagging groups were split into 2 tracer groups
each. Thus, the number of tagging groups not represented in any tracer
group, that is, the number of dispersed groups, is 88 or 22\% of the
number that were tagged. This fraction seems rather large and one
naturally wonders what happened to these dispersed groups.

The dispersed groups meet two characteristic fates involving the main
central halo. Groups that fall in along radial orbits are strongly
heated by interactions with the central potential of the cluster
and break up
completely. The particles diffuse in phase space such that no
identifiable core remains. Because of the formation process of groups
and clusters of galaxies in hierarchical models, radial orbits are
expected to be quite common (Katz \& White 1994, Summers, Evrard, \&
Davis 1995).
Other groups that orbit in the central potential are subjected to many
more smaller perturbations. Tidal effects cause these groups to
elongated along their orbits until their core is stretched apart. Note
that the split groups mentioned above are groups in an earlier stage
of being tidally stretched to disruption. Figure 9 provides examples
of both these types of dispersed groups.

The questions raised by these dispersed groups appear fatal to the
algorithm. One may argue that the strongly dispersed groups would be
expected to be parts of mergers and that, as long as a merger core
remains, not tracing these groups would not affect the population. In
practice, this would have to be done on a case by case basis and would
be time consuming. The more serious problem is that galaxy tracers
orbiting within a cluster seem destined to slowly diffuse and become
tidally stretched. This process is in the nature of the collisionless
simulation, and not just due to the most bound algorithm. Hence,
although the most bound algorithm is a significant improvement over the
halo and CC methods, one concludes that, in general, collisionless
galaxy tracers do not seem feasible if the individual galaxy halo
cores do not survive as sub-potentials within the cluster halo.

\section{SPH Galaxy Tracers}

The two fluid version of the simulation offers a marked advantage for
identifying galaxy tracers.  By including a baryonic species and the
relevant gas dynamical processes, one may follow the collapse process
past the halo stage.  Since the gas species is allowed to radiatively
cool, thermal pressure support is released and the gas collapses to
densities characteristic of the densities of real galaxies (Paper I).
Interpretation of the results should be much cleaner.

We note that the structures in the two fluid simulation, as defined by
the dark matter halos, are nearly identical to those of the single fluid
simulation.  Differences occur in the centers of halos where the
existence of a high density baryonic clump has made the dark matter more
concentrated. The details of the structure of the central cluster halo has
also been modified, but it will not affect the analysis here. No special
caveats are needed in comparing the two simulations.

With very high density contrasts in the baryons, picking out SPH galaxy
tracers is relatively straightforward. Briefly, we used the FOF grouping
algorithm with $\delta_{min}$ corresponding to roughly the observed
density of a present day galaxy and $N_{min}$ set to the resolution
limit of the SPH method (see Paper I for further details). Numerically,
these parameters at the final output are $\delta_{min}=8.7\xten{4}$ and
$N_{min}=30$ particles. For consistency, we will refer to these groups
of particles as SDE tracers instead of using the term galaxy-like
objects as in Paper I.

Plots of the SDE tracer distribution in the format of the above figures
is presented in Figure 10.  The baryons have collapsed to very high
overdensities and the tracers are very compact in the particle plots,
even though tens of thousands of points are plotted. In the field
regions, there is a good correspondence between halos and SDE tracers. A
few less SDE tracers are found than halos because, for systems at the
resolution limit, there is a time delay between when the halo forms and
when the gas inside it has cooled to the required density.  In the
central region, the reduced cross section of the baryonic tracers helps
them avoid the excessive merging of the halos and the inner few hundred
kpc is well populated. Being much like the halos, but resolving the
cluster region, these galaxy tracers fit all of the objective criteria
of \S 1.

SDE tracers satisfy basic plausibility criteria as well. The use of a gas
species that
can self consistently follow the collapse process greatly enhances one's
faith. The fact that complex identification schemes are unnecessary is
also a comfort. In addition, these tracers are the first which bear a
resemblance to real galaxy morphology and clustering (Summers 1993,
Paper I). The main argument against this method is that other physical
processes, such as star formation and associated feedback or ionizing
background radiation, are not included (see \S 2.2 of Paper I for a
detailed discussion). Unmodelled processes may significantly affect the
sites of galaxy formation or the evolution of the tracers. It is fair to
say that these concerns deal with evolving the state of the art and are
not gross omissions in the method. Future work will no doubt improve
the method.

Ours is not the only SPH simulation to find galaxy tracers of this sort.
In a similar but lower resolution study, KHW
employed essentially the same physics in a different code and also
formed very high density contrast objects. Their galaxy tracer
identification scheme was slightly different and a bit more lenient than
ours. To show the robustness of identifying SPH galaxy tracers in our
simulation, we investigated the KHW method as well. Following their
prescription, we used the SPH measured gas variables to identify all gas
particles at overdensities greater than 1000 and with temperature less
than 30{,}000 K. We then ran the FOF grouping algorithm on those
particles with $\delta_{min}=1000$ and $N_{min}=8$. Note that KHW used
the DENMAX procedure (Bertschinger \& Gelb 1991) as their grouping
algorithm, but, because of the high density contrasts involved, in only
a couple cases would this have produced a significant difference.

The KHW tracer population shows minor discrepancies with the SDE tracer
population. The lower overdensity cut allowed more tracer groups and
more particles in each tracer group into the population. The initial
overdensity filter, however, served to avoid spurious additions. These
extra groups and particles were not enough to significantly alter the
mass function of the tracers (presented in Figure 11). Figure 11 also
illustrates that the choice of $N_{min}=8$ is too lax for our
simulation. The integrated mass function flattens at about 30 particles
and indicates the resolution limit. KHW also recognized resolution
limitations and performed much of their analysis with $N_{min}=32$. We
shall use $N_{min}=30$ in section 4 below. Apart from these small
issues, both algorithms identified the same population of objects.

\section{Statistical Measures and Velocity Bias}

In the above sections, we tested the reliability of individual galaxy
tracer algorithms. Here, we examine the ramifications on
the inferred galaxy statistics under each method. Specifically, we
shall look at the classic statistical measures used in cosmology, the
correlation function and velocity dispersion of galaxies, under each
of the tracing methods. There does not exist a `correct' answer
against which the methods can be judged, but one can show the
variations in the statistics derived from the different populations.
Additionally, the results presented here are from a simulation of a
region too small to produce robust statistics and biased by the
constraint of forming a cluster of galaxies at the center. The numbers
should not be construed as general predictions of the CDM theory.

Five populations of galaxy tracers will be compared. As defined above,
they will be referred to as halos, CC, MB, SDE, and KHW. The
parameters of each algorithm have been set such as to define
approximately the same tracer population, i.e., those with a mass
greater than 30 particle masses. The exception is the CC method where
the lower mass cutoff is well defined at the tagging epoch but is
unknown at the final output because mass accretion is not considered.
The $N_{tag}$ and $N_{fin}$ parameters of the CC method were set as
generously as was prudent in order to produce a larger population. The
number of tracers for each method, in the order above, was 239, 108,
257, 215, and 238. The deficit in the CC method presumably reflects
the single tagging epoch: galaxy halos which form after $z_{tag}$ do
not make it into their population. As we naturally consider the SDE
algorithm to be the best of these tracer methods, we will use it as
the basis for comparison below, subject to the caveats of the previous
paragraph.

The correlation functions of the tracer populations are plotted in
Figure 12. As expected, SDE and KHW agree well. The differences below
75 kpc can be attributed to the uncertainty of small number
statistics.  The curves for both the halos and the MB tracers drop
under the SDE curve below 500 kpc. Such a drop in correlation is
expected for a method that does not adequately resolve clusters (Gelb \&
Bertschinger 1994). The
excess correlation seen in the CC method at all scales is a further
result of the single tagging epoch. Density enhancements that are
tagged early are the larger peaks of the density field and will
naturally aggregate over time and have large correlations (Kaiser
1984). The perturbations which collapse later and would dilute these
correlations are not included in the CC population. The agreement of
most methods on the largest scales indicates that collisionless
methods are adequate to estimate correlations on scales exceeding the
largest collapsed structures in the simulations.
Given the diversity of galaxy tracing algorithms, the broad agreement on
large scales may seem surprising. However, since the collapse is
pressureless for both fluids until a shock is reached, the baryons and
dark matter should track each other down to the collapsed scale.
Only the SPH methods
continue to follow the correlations on clustered scales. These results
reflect the transition from gravity dominated to hydrodynamics
dominated regimes.

The pairwise velocity dispersions of the tracer populations, shown in
Figure 13, do not provide a strong discriminant between the methods.
The dominant trend in the figure is that the SDE curve remains flat over a
large range of scales while the other methods are generally lower at
small scales and higher at large scales. The KHW curve differs only
below 100 kpc where the data are noisy.
At small
scales, one would attribute the SDE dispersion to its ability to resolve
the cluster center. The methods are converging to similar dispersions at
a few Mpc, but projection of this trend to agreement at larger scales
can not be verified without a larger dynamic range simulation.

Also plotted in Figure 13 is the pairwise velocity dispersion of the
dark matter particles. One sees immediately that all of the tracer
algorithms would predict a pairwise velocity bias from the simulation.
At a scale of 1 Mpc the ratio of tracer to dark matter velocity
dispersion ranges from 0.67 to 0.85. The SPH methods show the biggest
effect and the MB method has the least. Being optimistic, one might note
that the variation in estimates is not that large and agrees well with a
judiciously chosen sample from the literature. One could attribute the
range to varying amounts of friction, dynamical or viscous, and biased
evolution in the tracer populations. However, these estimates are
homogenized by using the same data set. When using both different tracer
methods and different simulations, the variance will increase strongly.

Further, this paper has shown that the tracer algorithms vary widely in
their reliability and one may fairly wonder whether any method has
measured a true velocity bias. For the halo algorithm, Gelb \&
Bertschinger (1994) pointed out that a reduced velocity dispersion is
expected when one removes the large velocity dispersions within a
cluster by representing them as a single halo. When they broke up the
large halos into a plausible galaxy distribution, the entire velocity
bias signal vanished.  This argument would also apply to the CC and MB
algorithms in that they are also underrepresenting the clusters.
Further, the MB results show that collisionless overdensities are tidally
destroyed in the
cluster region.  Thus dynamical friction would not be an important
effect for any of the collisionless tracer methods because the
overdensities don't remain tightly bound (a contrary opinion of
Carlberg (1994) will be discussed in \S5). As the MB method tracks
collisionless objects a bit better within the cluster, it is
consistent that it should show the least velocity bias. For the SPH
tracer methods, the very high overdensities achieved make them prime
targets for dynamical friction.  However, one can also argue that the
absence of star formation has kept the particles gaseous longer than
is reasonable and that ram pressure from the intracluster medium has
artificially slowed the galaxy tracers. In Paper I, we noted that the
stability of the velocity bias to both mass cuts and redshift of
observation (see Figure 26 of Paper I) argues against dynamical
friction, ram pressure and other secular effects. Here we mention that
another possibility is that a steady state is established, fed at
large radii by infall and drained at small radii by mergers. Such an
idea is motivated by numerical studies that include algorithms for
star formation and which show that ram pressure is an important effect
for purely gaseous SPH simulations (Summers 1993, Frenk \etal 1995).
For each method of tracing galaxies there exists questions as to whether
the velocity bias signal is enhanced or even created by charcteristics
of the tracing algorithm.

We conclude that definitive predictions on the existence of velocity
bias as a general phenomenon awaits further development of galaxy tracing
within simulations. Couchman and Carlberg (1992) based their claims on a
poor galaxy tracer algorithm and the statistics of their tracer
population are not reliable. Carlberg (1994) utilizes much higher
dynamic range, but still uses a similar algorithm that does not solve
the basic problems with collisionless tracer methods.  The velocity bias
described in Paper I may be overstated due to ram pressure effects in
the cluster.  Our previous result and the lack of velocity bias seen by
KHW are from simulated regions that are too small for general
statistics. The data presented by Lubin and Bahcall (1993, see their
Figure 5) indicate that the velocity bias parameters may show factor of
2--4 variation from cluster to cluster.
Lacking wide dynamic range simulations that model all of the relevant
physical processes, the existence and magnitude of velocity bias remains
an open question.

\section{Discussion}

This paper has examined many of the most popular methods for tracing
galaxies in cosmological simulations. A brief summary of the results
will help us to identify future improvements in the methods. One must
also recognize the limitations in using simulations to test theories of
structure formation and special attention will be paid to the
contentious issue of velocity bias. From this discussion, a clear picture
of the current state of affairs should emerge.

\subsection{Summary}

In evaluating the basic methodology, four algorithms which identify
galaxy tracers in collisionless simulations were considered. The
standard methods of identifying dark matter halos from the evolved
density field or marking peaks in the initial density field have
previously documented deficiencies: the halo method does not resolve
cluster regions well and the initial peak tracers show insufficient
correlation with the non-linear structures. The algorithm of Couchman \&
Carlberg (1992) is well motivated, but ill conceived in that it fails to
address coherence, continuous formation, and growth of tracers. The most
bound algorithm is developed here to circumvent the above problems, but
finds that clusters are still poorly resolved due to heating and
tidal disruption of tracers. Diffusion in phase spase of a collisionless
fluid may be an insurmountable problem.

The addition of smoothed particle hydrodynamics markedly improves the
situation. Dissipation produces very high density contrasts that can be
traced directly. Minor differences in the results from the methods of
Evrard \etal (1994) and Katz \etal (1992) do not favor either algorithm.
Rather, the limitations now arise in the physical model of the
simulation code, not in the post-processing tracer identification. The
addition of a baryonic species opens up questions about the relative
roles of processes like ram pressure, radiative cooling, star formation,
and supernova in the collapse of a galactic scale perturbation.

Ultimately, one wants to get robust statistical measures from the galaxy
tracer algorithms. In a group comparison of the estimated correlation
functions, most methods agree at large scales and the methods that do
not adequately resolve the cluster show a decline at small
separations. The CC method finds excess correlation strength at all
scales due to biasing inherent in the algorithm. SPH methods produce
the most consistently behaved estimate, as is true for the velocity
dispersions as well. The small scale drop in signal is also present in
velocity dispersions, but convergence of opinion occurs only on the
largest scale addressed here (few Mpc). This variance in statistical
estimates by simply changing the tracer method signals the need for
caution and careful interpretation of simulation results.

\subsection{Improvements to Tracer Methods}

In future work on collisionless galaxy tracers, the question of the
coherence of halos orbiting in a cluster potential will have to be
settled. Carlberg (1994) has suggested that increased dynamic range in
mass and length scales will remove the problem of two body heating and
allow orbits to be traced internal to the cluster. His estimates do not
consider the tidal effects due to the cluster and due to other orbiting
halos. Work by Moore \etal (1995) finds that such tidal
effects can become the dominant source of heating and lead to dispersal,
depending on the assumed density profile of the halos. As there is not a
consensus on the infinite resolution limit of collisionless halo density
profiles, all would agree that the dissolution of substructure must be
probed further.

Other progress on collisionless methods will involve a shift of
paradigm. Instead of utilizing post-processing techniques, one can
incorporate galaxy tracing ideas into the simulation code.  One idea
that has been tried is to use ``sticky'' particles, i.e., particles
which undergo inelastic collisions (Carlberg 1988). This method is a
crude way to mimic hydrodynamics, but, in our unpublished tests, does
not seem to offer much advantage over SPH simulations in CPU time.
Alternatively, one might use massless tracers in a most bound algorithm
to follow the evolution of the center of the potential well. Like the
peaks method, problems here involve following the merging or disruption
of these tracers and dealing with individual particle tidal scattering.
Another idea is to introduce `artificial cooling' into a collisionless
simulation by collecting particles in a collapsing region into a more
massive superparticle. This method is being studied by van Kampen
(1995),
however one must be careful to avoid spurious numerical effects due to
a large mass ratio between particles in a simulation (Peebles \etal
1989, this caution also applies to SPH simulations, although to a lesser
degree because hydrodynamic forces tend to dominate gravity in dense
regions).
All of these ideas have the highly undesirable feature that one
has to re-run the simulation in order to change parameters of the galaxy
tracing method.

Collisionless simulations will always lack the hydrodynamic processes
that determine much of the evolution of the luminous parts of
galaxies.  For SPH models, the challenge is to investigate the effects
of including further physics. An ionizing background radiation at high
redshift may suppress the cooling and delay or prevent galactic
collapse. Handling star formation has several associated concerns:
keeping the particles gaseous can overstate ram pressure effects,
transforming gas particles to a collisionless stellar fluid with
limited resolution may resurrect phase space diffusion problems, and
modelling of supernova energy feedback can effectively control the
galaxy formation rate (Navarro \& White 1993). Further issues of
tracking metallicity and the related cooling of atomic and molecular
species will keep the field busy for quite some time.  Another
limitation is imposed by the mass resolution required for galaxy sized
clouds to cool efficiently.  The runs described herein are the largest
SPH simulations in this area of research and can trace galaxies well
in simulations up to about 25 Mpc.  This scale is a bit small for
determining global statistics of cosmologies. The route past this
block will be to perform ensemble sets of these simulations or to
utilize parallel techniques to produce larger simulations.  The use of
SPH in cosmological simulations is also a relatively new field and
testing and refining of the algorithm will continue.

The research avenues are not limited to those discussed above. Eulerian
(grid based) hydrodynamic codes have been applied to cosmology, but do
not currently have sufficient dynamic range to address galaxy formation
in other than a heuristic fashion (Cen \& Ostriker 1992). Methods of
softened LaGrangian hydrodynamics (Gnedin 1995) and moving mesh
algorithms (Pen 1995) can add about a factor of ten improvement to grid
techniques. Much wider range will be achieved by three dimensional
adaptive grid methods under development (Neeman 1994).
The multi-pronged attack on the problem offers a wide
base for success.

\subsection{Interpretation of Tracer Statistics}

These improved simulations will best serve as predictions of
cosmological models when analyzed carefully. This paper has shown that
the influence of galaxy tracing methods must be recognized and stated
explicitly. The range of validity of the derived results should be
estimated and, where possible, the comparison of several methods can be
quite illuminating. To be specific, the correlations of galaxies seem
quite robust down to the virialized scale of clusters while velocities
are stable only to a scale a few times larger. It would be desirable to have
larger simulations which could confirm that these relationships hold into the
linear regime.

In general, the results from computer simulations can be viewed in a manner
similar to observational data. One is quite comfortable taking into
account such things as point spread functions, sky subtraction, smoothing,
and model fitting when processing data, and the same types of ideas apply
to numerical work. Both author and reader may need to be more careful in
noting these details.

A case in point is the discussion of velocity bias. Estimates in the
literature span a factor of 3 in value. Different groups disagree about
dependence on separation scale, mass of the galaxy population, and
observation epoch. Only recently has the distinction between pairwise
velocity bias and cluster velocity bias been explicitly stated. We
suggest here that much of this variation is due to different galaxy
tracing techniques. Even if all tracer methods produced the exact same
estimate of velocity bias, the errors in each estimate due to the
inadequacies of the individual methods would still demand substantial
error bars.  The errors introduced by the tracer algorithm
typically have not been stated explicitly.

Robust estimates will be a computational challenge. The galaxy formation
process must be followed through three stages: pressureless initial
perturbation growth, dissipational gas collapse, and establishment of a
collisionless stellar fluid.  To cover scales from sub-galactic to a
fair sample of the universe requires dynamic ranges of order $10^4$ in
length and $10^7$ in mass. Cosmic variance may require an ensemble of
simulations to pin down a value for a particular model. Plus, currently
there are several cosmological models from which to choose.

On the theoretical side, the nature of velocity bias has yet to be
agreed upon. In the absence of damping on short timescales, the
velocities of galaxies must reflect the depth of the potential well,
however, it is not necessary that galaxies have the same radial
distribution as the mass. Those who believe that galaxies have a
biased spatial distribution, must accept the corollary that the
velocities will also be biased. One may distinguish between the
probable sources, dynamical friction, peak biasing, or natural
biasing, by isolating their effects in controlled tests. Additionally,
a point that is often lost in the noise is that velocity bias is not
an end in itself, but is most relevant when applied to clusters of
galaxies and their mass estimates.  A biased spatial and velocity
distribution implies that the standard virial mass estimation may not
cover the full virial radius and may not reflect the total virial mass
of the cluster.  The increasing data on mass estimates from velocity
dispersions, x-ray temperature profiles, and gravitational lensing
will allow one to compare different estimates at the same radii and
get some measure of the effect in the real universe. Several methods
of estimation are important for helping constrain such unknowns as
projection effects and mass profile models.  An understanding of
velocity bias is an important step in characterizing the dynamical
evolution of galaxies.

\subsection{Conclusions}

Cosmological simulations are beginning to be able to do what one wanted
them to do a decade ago: provide realistic and detailed predictions of
the distribution of galaxies that form in a theory. They are a useful,
but quite complex tool. A cautious and studied approach is necessary to
gain the confidence of the astronomical community in the results.

Galaxy tracers in numerical models can play a critical role in
determining the predictions of a simulation. Though much emphasis has
heretofore been placed on the evolution codes, considerable focus must
also be put on the analysis procedures.  None of the galaxy tracing
algorithms examined here is perfect. Each algorithm has constraints
that must be identified and the ensuing limitations on the results
should be specified.  The methods of tracing galaxies need to be
examined before being implemented: theoretical plausibility does not
guarantee practical utility.  The future of this field appears to be a
strong one, but it will necessarily proceed slowly and steadily.

\bigskip

The authors wish to thank R. Carlberg, N. Katz, and D. Weinberg for
helpful discussions. Support for this work was provided by NSF grant
AST-8915633 and NASA grant NAGW-2367. Computing resources provided by the
San Diego Supercomputing Center and the Center for Particle Astrophysics
are gratefully acknowledged.

%
%

\startreferences

\jref
Bardeen, J. M., Bond, J. R., Kaiser, N., \& Szalay, A. S.;1986;\apj;304;15;.

\jref
Barnes, J.;1989;\nature;338;123;.

\jref
Bertschinger, E., \& Gelb, J. M.;1991;\cip;Mar/Apr;164;179.

\jref
Bertschinger, E., \& Jain, B.;1994;\apj;431;486;494.

\jref
Carlberg, R. G.;1988;\apj;324;664;.

\jref
Carlberg, R. G.;1994;\apj;433;468;.

\jref
Carlberg, R. G., \& Couchman, H. M. P.;1989;\apj;340;47;.

\jref
Carlberg, R. G., Couchman, H. M. P., \& Thomas, P. A.;1990;\apjl;352;L29;L32.

\jref
Carlberg, R. G., \& Dubinski, J.;1991;\apj;369;13;.

\jref
Cen, R., \& Ostriker, J.;1992;\apjl;399;L113;L116.

\jref
Couchman, H. M. P., \& Carlberg, R. G.;1992;\apj;389;453;463.

\jref
Davis, M., Efstathiou, G., Frenk, C. S., \& White, S. D. M.;1985;\apj
;292;371;394.

\jref
Davis, M., \& Peebles, P. J. E.;1983;\apj;267;465;.

\jref
Davis, M., Summers, F J, \& Schlegel, D.;1992;\nature;359;393;396.

\jref
Efstathiou, G., Bond, J. R., \& White, S. D. M.;1992;\mnras;258;1p;6p.

\jref
Efstathiou, G., Sutherland, W. J., \& Maddox, S. J.;1990;\nature
;348;705;707.

\jref
Evrard, A. E.;1988;\mnras;235;911;.

\jrefabbrev
Evrard, A. E., Summers, F J, \& Davis, M.;1994;\apj;422;11;36;Paper I.

\jrefsub
Fisher, K., Strauss, M., \& Davis, M.;1995;\apj;;;.

\jrefsub
Frenk, C. S., Evrard, A. E., White, S. D. M., \& Summers,
F J;1995;\apj;;;.

\jref
Gelb, J. M. \& Bertschinger, E.;1994;\apj;436;491;508.

\jrefpress
Gnedin, N. Y.;1995;\apjs;;;.

\jref
Katz, N., Hernquist, L., \& Weinberg, D. H.;1992;\apjl;399;L109;L112.

\jref
Katz, N., Quinn, T., \& Gelb, J. M.;1993;\mnras;265;689;705.

\jref
Katz, N. \& White, S. D. M.;1994;\apj;412;455;478.

\jref
Kaiser, N.;1984;\apjl;284;L9;L12.

\jref
Klypin, A., Holtzman, J., Primack, J., \& Reg\"os, E.;1993;\apj;416;1;16.

\jref
Lubin, L. M., \& Bahcall, N. A.;1993;\apjl;415;L17;L20.

\jref
Melott, A. L., Einasto, J., Saar, E., Suisalu, I., Klypin, A. A.,
\& Shandarin, S. F.;1983;\physrevlett;51;935;938.

\jrefpress
Moore, B., Katz, N., \& Lake, G.;1995;\apjl;;;.

\jref
Navarro, J. F., \& White, S. D. M.;1993;\mnras;265;271;300.

\jrefprivcomm
Neeman, H.;1994;;;;.

\jref
Park, C.;1991;\mnras;251;167;173.

\jref
Peebles, P. J. E., Melott, A. L., Holmes, M. R., \&  Jiang, L. R.;1989;
\apj;345;108;121.

\jrefinprep
Pen, U.;1995;\apjs;;;.

\bref
Summers, F J 1993, Ph{.}D. Thesis, University of California at
Berkeley.

\jrefinprep
Summers, F J, Evrard, A. E., \& Davis, M.;1995;\apj;;;.

\jrefsub
van Kampen, E.;1995;\mnras;;;.

\jref
Vogeley, M. S., Park, C., Geller, M. J., \& Huchra, J. P.;1992;\apjl
;391;L5;L8.

\jref
White, S. D. M., Frenk, C. S., \& Davis, M.;1983;\apjl;274;L1;L5.

\jref
White, S. D. M., Frenk, C. S., Davis, M., \& Efstathiou, G.;1987;\apj
;313;505;516.

\endreferences

%
%

\startfigcapt

\figcapt{1}
Two plots which characterize the dark matter distribution in the
simulation. Length scales here and throughout are given in physical units at
$z=1$. On the left is the entire simulation volume (a 7 Mpc cube)
showing the central group, a second group above and right of center, and
various filamentary structures. For clarity, only one fourth of the
particles are plotted. The right hand side details the central region,
one tenth the box length on a side, and shows all of the particles.

\figcapt{2}
Two plots of the groups found by the halo algorithm. Format is similar
to Figure 1. The left hand plot shows the halos in the entire
simulation volume. Note that some halos do not appear distinctly in
Figure 1 because only one fourth of the particles are plotted in that
figure. The right hand side details the halos in the central region (one
tenth the box size) with circles denoting the center of mass positions
of the smaller halos. The dominant halo is roughly 400 kpc across and
should contain many galaxies.

\figcapt{3}
An example of the merging of dark matter halos. Both panels are cubic
regions three tenths of the box length on a side and show the particles
that wind up in the largest halo at the final output.  The left hand
plot shows the large amount of structure that was evident at $z=2$
and is erased by the final output, $z=1$ (shown on the right).

\figcapt{4}
CC galaxy tracers found when $\delta_{tag}$ is used to set a constant
physical density criterion. The left panel, at $z_{tag}=6.6$, shows the
particles in the entire box which are tagged by the first grouping
algorithm.  The right panel, also the entire box, but now at
$z_{fin}=1.0$, shows the particles which pass through the second
grouping cut in the CC method and would be classified as galaxy tracers.
For clarity, circles have been drawn around the center of mass of the
tracers in the right hand panel which lie outside the central region.

\figcapt{5}
CC galaxy tracers found when $\delta_{fin}$ is used to set a
constant physical density criterion. Format is the same as Figure 4,
except that no circles are drawn in the right hand panel.

\figcapt{6}
CC galaxy tracers found when setting $\delta_{tag}$ and
$\delta_{fin}$ independently. Format is the same as Figure 2.

\figcapt{7}
Effects of a mass range cut on the CC tracers in the very central
region. These plots show the CC galaxy tracers in a cubic region with
side length one twentieth of the box length (350 kpc, one half that of
the right plot in Figure 6). The left panel shows the largest tracer,
which would be excluded from a tracer population by a mass range cut.
The right panel shows the tracers that remain after the largest one is
removed from the region.

\figcapt{8}
Galaxy tracers found with the Most Bound algorithm. Format is the same
as Figure 2.

\figcapt{9}
Tidal interactions on MB galaxy tracers. Shown here are two groups that
were tagged by the MB algorithm, but which have become tidally dispersed
by the end of the simulation. Both panels show the same 700 kpc cubic
region as the right panel of Figure 8, but from different projections.
Velocities are relative to the center of mass velocity and are plotted
as tails. The left panel gives an example of a tracer that has been
tidally extended along its orbit through the central halo. The tracer in
the right panel has been scattered throughout the region by interactions
with the central potential.

\figcapt{10}
Galaxy tracers found using the algorithm of SDE. Format of the plots is
the same as Figure 2, but the central region is shown in a different
projection.

\figcapt{11}
Integrated mass function of the SDE and KHW galaxy tracers.  The number
of tracers greater than a given mass is plotted versus mass down to the
resolution limit of the methods. Mass here refers to the baryonic mass
in the tracers.

\figcapt{12}
Correlation functions for the galaxy tracing methods discussed in
section 4. The comoving length scales are twice the physical scales
given in previous figures because the simulation was stopped at $z=1$.

\figcapt{13}
Pairwise velocity dispersions of the galaxy tracers for the methods
discussed in section 4. Also plotted is the pairwise velocity
dispersion of the dark matter particles, i.e., the velocity dispersion
of the dominant mass.

\endfigcapt

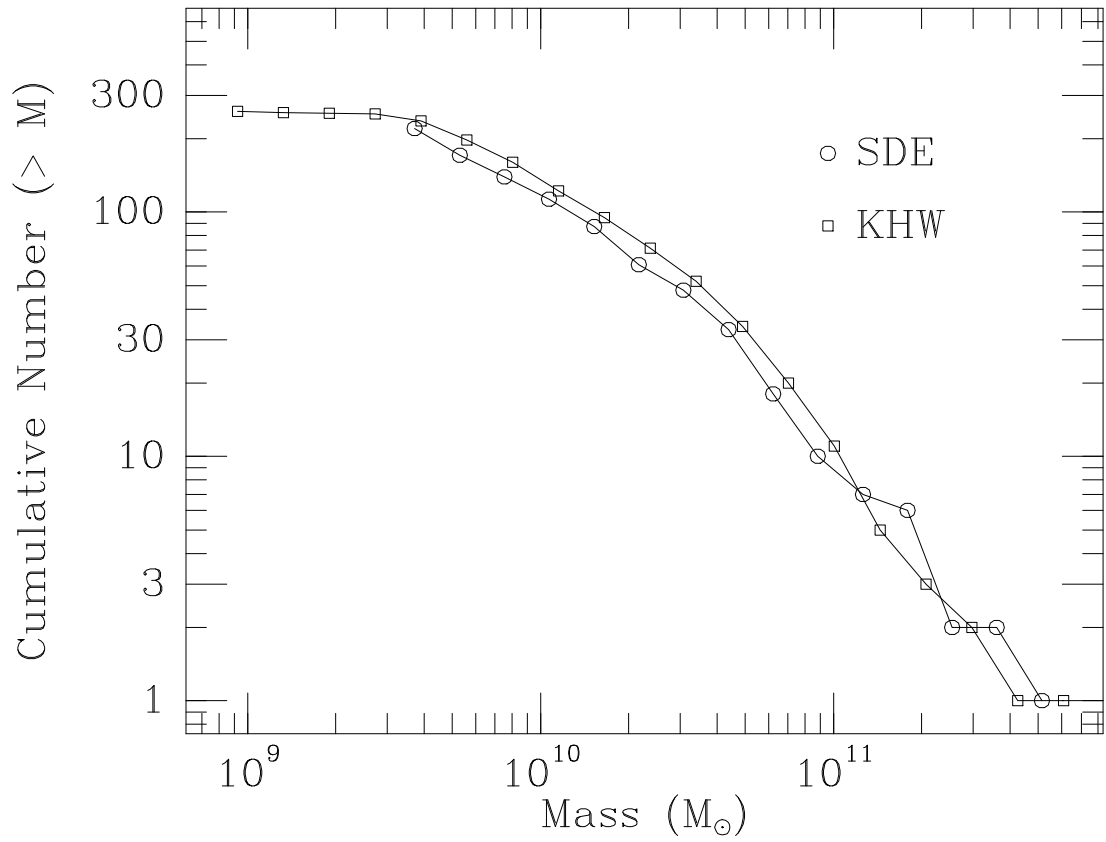
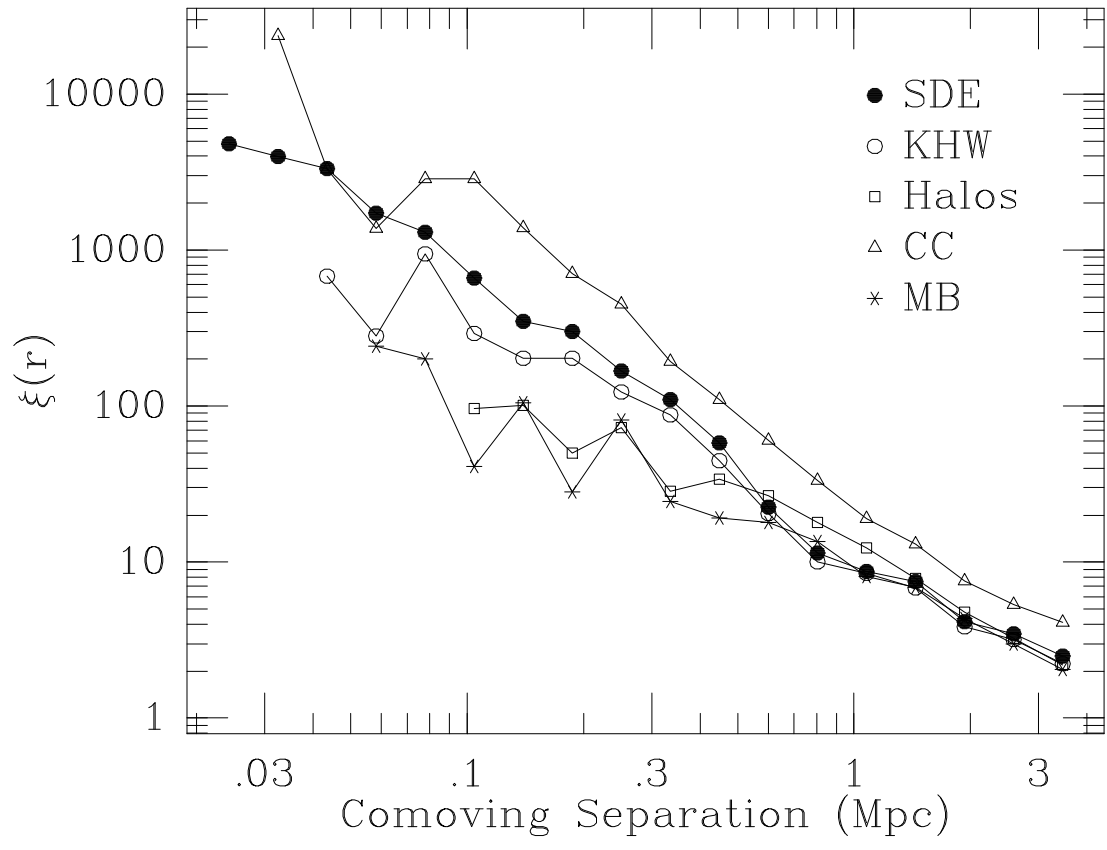
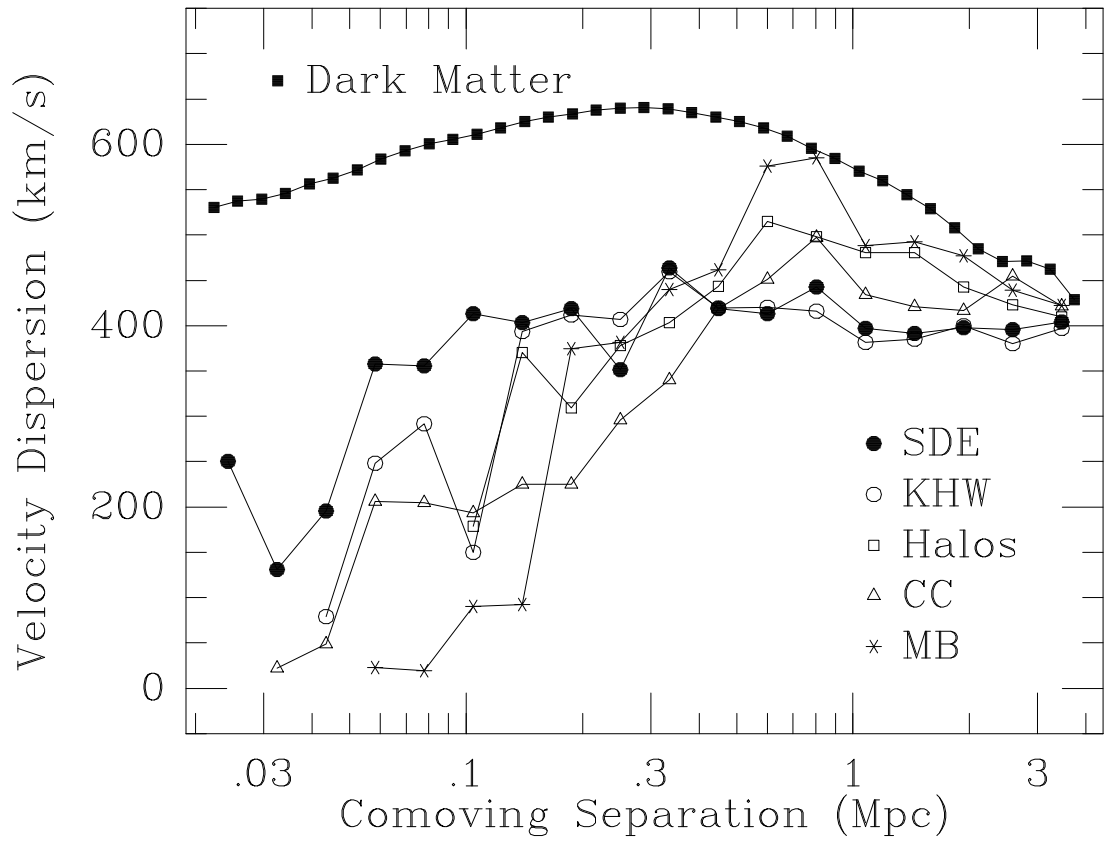
\end